\documentclass{article}
\usepackage{LaThuileFPSpro}
\usepackage{xspace}                                 
\begin{document}
\newcommand{\abs}[1]{\left| #1 \right|}

\newcommand{\xs}{\ensuremath{\sigma_{t\overline{t}}}}
\newcommand{\mt}{\mbox{M\ensuremath$_{\mbox{top}}$}}
\newcommand{\mtr}{m\ensuremath{_t^{\mbox{reco}}} }
\newcommand{\wjj}{m\ensuremath{_{\mbox{jj}}} }
\newcommand{\Mvariable}[1]{#1}
\newcommand{\mtnwa}{M\ensuremath{_t^{\mbox{NWA}}}\xspace}

\newcommand{\instlumunits}[1] {\ensuremath{#1\times 10^{30}\mbox{cm}^{-2}\mbox{sec}^{-1}}\xspace}
\newcommand{\hminus} {\ensuremath{\mathrm{H}^{-}}\xspace}
\newcommand{\pbar}  {\ensuremath{\bar{p}}\xspace}
\newcommand{\ppbar}  {\ensuremath{p\bar{p}}\xspace}
\newcommand{\ttbar}  {\ensuremath{t\bar{t}}\xspace}
\newcommand{\qqbar}  {\ensuremath{q\bar{q}}\xspace}
\newcommand{\bbbar}  {\ensuremath{b\bar{b}}\xspace}
\newcommand{\ccbar}  {\ensuremath{c\bar{c}}\xspace}
\newcommand{\zll}    {\ensuremath{Z\rightarrow l^{+}l^{-}}}

\newcommand{\pte}    {\ensuremath{p_{T}^{e}}}
\newcommand{\ptmu}   {\ensuremath{p_{T}^{\mu}}}
\newcommand{\ptnu}   {\ensuremath{p_{T}^{\nu}}}
\newcommand{\ptjet}  {\ensuremath{p_{T}^{jet}}}
\newcommand{\ptpart} {\ensuremath{p_{T}^{parton}}}
\newcommand{\ptw}    {\ensuremath{p_{T}^{W}}}
\newcommand{\ptz}    {\ensuremath{p_{T}^{Z}}}

\newcommand{\etadet} {\ensuremath{\eta_{det}}}
\newcommand{\etajet} {\ensuremath{\eta^{jet}}}
\newcommand{\etjet}  {\ensuremath{E_{T}^{jet}}}

\newcommand{\Ht}     {\ensuremath{H_{T}}\xspace}
\newcommand{\et}     {\ensuremath{E_{T}}}
\newcommand{\ex}     {\ensuremath{E_{x}}}
\newcommand{\ey}     {\ensuremath{E_{y}}}
\newcommand{\pt}     {\ensuremath{p_{T}}}
\newcommand{\px}     {\ensuremath{p_{x}}\xspace}
\newcommand{\py}     {\ensuremath{p_{y}}\xspace}
\newcommand{\pz}     {\ensuremath{p_{z}}\xspace}
\newcommand{\met}    {\mbox{$\protect \raisebox{.3ex}{$\not$}\et$}\xspace}
\newcommand{\mex}    {\mbox{$\protect \raisebox{.3ex}{$\not$}\ex$}}
\newcommand{\mey}    {\mbox{$\protect \raisebox{.3ex}{$\not$}\ey$}}

\newcommand{\chisq}  {\ensuremath{\chi^{2}}\xspace}
\newcommand{\chisqmin}{\ensuremath{\chi^{2}_{\mathrm{min}}}\xspace}
\newcommand{\mtop}   {\ensuremath{\mathrm{M}_{\mathrm{top}}} }
\newcommand{\runi}   {run~I\xspace}
\newcommand{\runii}  {run~II\xspace}

\newcommand{\genunit}[2]{\ensuremath{#1~\mathrm{#2}}\xspace}
\newcommand{\degrees}[1]{\ensuremath{#1^{\mathrm{o}}}\xspace}
\newcommand{\meters}[1]{\ensuremath{#1~\mathrm{m}}}
\newcommand{\kev}[1] {\ensuremath{#1~\mathrm{keV}}}
\newcommand{\mev}[1] {\ensuremath{#1~\mathrm{MeV}}}
\newcommand{\mevcc}[1] {\ensuremath{#1~\mathrm{MeV}/c^{2}}}
\newcommand{\tev}[1] {\ensuremath{#1~\mathrm{TeV}}}
\newcommand{\gev}[1] {\ensuremath{#1~\mathrm{GeV}}}
\newcommand{\gevnoarg}{\ensuremath{\mathrm{GeV}}}
\newcommand{\gevc}[1] {\ensuremath{#1~\mathrm{GeV}/c}}
\newcommand{\gevcnoarg}{\ensuremath{\mathrm{GeV}/c}}
\newcommand{\gevcc}[1]{\ensuremath{#1~\mathrm{GeV}/c^{2}}}
\newcommand{\gevccnoarg}{\ensuremath{\mathrm{GeV}/c^{2}}}
\newcommand{\invpb}[1]{\ensuremath{#1~\mathrm{pb}^{-1}}}
\newcommand{\invfb}[1]{\ensuremath{#1~\mathrm{fb}^{-1}}}
\newcommand{\invpbnoarg}{\ensuremath{\mathrm{pb}^{-1}}}

\newcommand{\scinot}[2]{\ensuremath{#1\times 10^{#2}}\xspace}
\newcommand{\djes}{\ensuremath{\Delta_{\mbox{JES}}}\xspace}
\newcommand{\sigmac}{\ensuremath{\sigma_{\mbox{c}}}\xspace}

\newcommand{\measErr}[2]{\ensuremath{#1 \pm #2}}
\newcommand{\measAErr}[3]{\ensuremath{#1~^{+#2}_{-#3}}}
\newcommand{\measStat}[2]{\ensuremath{#1 \pm #2~\mathrm{(stat.)}}}
\newcommand{\measAStat}[3]{\ensuremath{#1~^{+#2}_{-#3}~\mathrm{(stat.)}}}
\newcommand{\measStatSyst}[3]{\ensuremath{#1 \pm #2~\mathrm{(stat.)} \pm #3~\mathrm{(syst.)}}}
\newcommand{\measAStatSyst}[4]{\ensuremath{#1~^{+#2}_{-#3}~\mathrm{(stat.)}\pm #4~\mathrm{(syst.)}}}
\newcommand{\measStatSystDZero}[5]{\ensuremath{#1~^{+#2}_{-#3}~\mathrm{(stat+syst.)}~^{+#4}_{-#5}~\mathrm{theory}}}

\title{ 
  TOP QUARK MASS MEASUREMENTS AT THE TEVATRON
  }
\author{
  Jahred Adelman \\
  (on behalf of the CDF and D\O\ collaborations) \\
  {\em University of Chicago. 5640 S. Ellis Avenue, Chicago, IL 60615, USA}
  }
\maketitle

\baselineskip=11.6pt

\begin{abstract}
Top quark mass measurements from the Tevatron using up to \invfb{2.0} of data are presented. Prospects for combined Tevatron measurements by the end of Run II are discussed.
\end{abstract}
\newpage
\section{\label{sec:Intro}Introduction}
Discovered in 1995 by both CDF and D\O, the top quark is by far the heaviest known fundamental particle \cite{CDFdiscovery, DZerodiscovery}. The mass of the top quark (\mtop) is of particular interest, as radiative contributions involving both the top quark and the putative Higgs boson contribute to the mass of the W boson. Thus, the masses of the top quark, the Higgs boson and the W boson are not three independent parameters in the Standard Model (SM). When and if the Higgs boson is discovered, precision measurements of the masses of the W boson and the top quark will help make a key test of the SM, helping to answer whether the new find is indeed the SM Higgs boson or some other, new scalar particle. In addition, the heavy mass of the top quark, near the electroweak scale, indicates that the top quark may play a role in helping theorists disentangle possible new sources of physics \cite{newphysics}. This letter describes measurements of the top quark mass from the CDF and D\O\  collaborations using up to \invfb{2.0} of data collected in Run II at the Tevatron.

\section{\label{sec:hard}Production and Decay}
 Top quarks at the Tevatron are produced predominantly in pairs, and decay almost always in the SM to a W boson and a b quark. The topology of \ttbar events depends on the subsequent decay of the two W bosons. In the dilepton channel, each W boson decays leptonically, to an electron or muon and a neutrino. The dilepton channel has the lowest background and only two jets in the leading order \ttbar decay, but suffers from underconstrained kinematics due to the two escaping neutrinos, as well as from having the lowest branching fraction among all decay channels. In the all-hadronic channel, the two W bosons decay hadronically to quarks. The all-hadronic channel has the largest branching fraction and no neutrinos, but also contains no charged lepton to distinguish it from the large QCD background. In the lepton+channel channel, one W boson decays hadronically and the other leptonically. Though there is an undetected neutrino, the kinematics of the system are still overconstrained.

Two tricks are used often in \ttbar mass analyses to increase the signal-to-background and improve systematics. Each \ttbar event contains two btags; if the secondary vertices from the decay of metastable B hadrons can be identified, jets arising from b quarks can be distinguished from jets arising from light flavor quarks. This significantly cuts down on the number of background events, and also helps to match the jets observed in the detector to the quarks at the hard scatter level. Lepton+Jets and all-hadronic events also contain at least one hadronically decaying W boson. The narrow decay width and well known W boson mass in these events can be used to constrain, {\it in situ}, the largest systematic in top quark mass measurements, the calibration and response of calorimeters to hadronic particles, also known as the jet energy scale (JES).

\section{\label{sec:dileptons}Dilepton template analyses}
Due to the underconstrained kinematics, measurements of \mtop in the dilepton channel must integrate over some unknown quantities. The D\O\ experiment has two dilepton measurements, each using \invfb{1} of data. In the matrix weighting method, each charged lepton-jet pairing is given a weight for the expectation to find, within experimental resolutions, the leptons with the measured energy, given a top quark mass and the unknown top and and anti-top \pt. The \pt\ values are integrated over using parton distribution functions, and a likelihood fit yields \mtop = \gevcc{\measStatSyst{175.2}{6.1}{3.4}} \cite{d0matrixweighting}, with systematics that are dominated by the jet energy scale. In the Neutrino Weighting Algorithm (NWA), the unknown pseudorapidities of the two neutrinos are integrated over. The solutions for a given top quark mass are weighted by the agreement with the missing transverse energy in the detector. The mean and RMS of the top quark mass weight distribution are used as estimators for the true top quark mass. With \invfb{1} of data, D\O\  measures \gevcc{\measStatSyst{172.5}{5.8}{3.5}} \cite{d0nwa}. The above two measurements, while largely correlated, are not completely correlated. A combination using the BLUE technique \cite{blue} yields \gevcc{\measStatSyst{173.7}{5.4}{3.4}} \cite{d0dilcomb}.

CDF uses the NWA measurement in the dilepton channel with \invfb{1.9} of data. The most probable top quark mass, and not the mean, is taken as the first estimator; the distribution is often rather asymmetric, so these are not necssarily the same quantity. The second observable is the \Ht, the scalar sum of \met, lepton \pt\ values and jet \pt\ values. CDF measures \gevcc{\measAStatSyst{171.6}{3.4}{3.2}{3.8}} \cite{cdfUS}.

\section{\label{sec:lepton+jets}Other template analyses}
The kinematics in the lepton+jets and dilepton channel are overconstrained, so there is no need to integrate over unknown quantities. The overconstrained kinematics are also used to select the single best assignment of jets to the quarks at the hard scatter--the single assignment most consistent with the \ttbar hypothesis is used. CDF has two such measurements with \invfb{1.9} of data. A measurement in the all-hadronic channel uses a neural network to increase the S:B and reduce the QCD background. In addition, the W mass constraint is used to calibrate the JES, yielding \gevcc{\measStatSyst{177.0}{3.7}{1.6}} \cite{cdfallhad}, where, as in all such measurements that contain an {\it in situ} JES calibration, the statistical uncertainty also includes a component for the JES systematic that now scales with $1/\sqrt{N}$. A measurement in the lepton+jets channel yields \gevcc{\measStatSyst{171.8}{1.9}{1.0}} \cite{cdfUS}, and also includes an {\it in situ} calibration. CDF also has the first-ever analysis combining measurements of the top quark mass across different decay toplogies into the same likelihood. More-traditional combinations must assume correlations for systematics between measurements, as well as assume Gaussian behavior of the separate likelihoods. By combining the measurements into the same likelihood, these assumptions are not needed. The combination of CDF's lepton+jets and dilepton anaylses described above yields \gevcc{\measStatSyst{171.9}{1.7}{1.0}} \cite{cdfUS}.

\section{\label{sec:me}Matrix Element Analyses}
A different class of top quark mass analyses, called matrix element (ME) analyses, try to extract as much information as possible from every event. All jet-parton assignments consistent with b-tagging are used in the likelihood, which makes use of leading order theoretical predictions for how \ttbar events are produced and decayed, as given by the matrix element. Typically, leptons are assumed to be perfectly measured, as are jet angles. The energies of the partons at the hard scatter level are encoded in transfer functions, which give the probability to observe a jet with energy $j$ given a parton with energy $p$. The transfer functions are needed since analyses measure jets in the detector, but the matrix element knows only how to describe events at the parton level. In typical ME analyses, the probability to observe $\vec{x}$ in the detector, given some top quark mass and JES in the detector, is given by:

\begin{equation}
P(\vec{x} | \mtop, \mbox{JES}) = \frac{1}{N}\int d\Phi | M_{\ttbar}(p; \mtop)|^2 \prod_{\mbox{objects}}W(j|p,\mbox{JES})f_{\mbox{PDF}}(q1,q2),
\end{equation}
\noindent
where P gives the probability to observe x in the detector, given some top quark mass (and JES in the detector, if the measurement includes an {\it in situ} calibration). N is a normalization term that includes effects of efficiency and acceptance, as well as the changing \ttbar production cross section as a function of \mtop. The integral over $d\Phi$ is an integral over the parton-level phase space. The matrix element M is the leading order matrix element for \ttbar production with partons $p$, given some top quark mass. The transfer functions W give the probability to observe a jet with energy $j$ given a parton energy $p$ (and possibly the jet calibration in the detector). Finally, there are two terms in $f_{\mbox{PDF}}$ that come from the parton distribution functions and give the probability to observe the two incoming partons with the appropriate energy.

CDF has a ME element in the dilepton channel using \invfb{2.0}. The analysis uses a novel evolutionary neural network at the selection stage to improve the {\it a priori} statistical uncertainty on the top quark mass by 20\%. Normal neural networks are trained only to minimize misclassifcation. As such, they can be used only to distinguish signal and background, not to improve directly the expected uncertainty on a measurement. The analysis measures \mtop = \gevcc{\measStatSyst{171.2}{2.7}{2.9}} \cite{cdfmadcow}.

CDF has a ME element analysis in the lepton+jets channel using \invfb{1.9}. The analysis differs from typical ME analyses via the modification of the propagators in the matrix element to account for the imperfect assumptions about perfectly measured angles and intermediate particle masses that make the multi-dimensional integral tractable. The analysis also makes a cut on the peak likelihood to remove both background events as well as poorly modeled signal events where the object in the detector do not match the assumed partons at the matrix element level. The analysis includes an {\it in situ} JES calibration, and measures \mtop = \gevcc{\measStatSyst{172.7}{1.8}{1.2}} \cite{cdfmeat}. D\O\  also has a ME analysis using \invfb{0.9}. Unlike most lepton+jet analyses, this analysis includes events with 0 b-tags. The events are separated by charged lepton type (electron vs muon). Including an {\it in situ} JES calibration, D\O\  measures \mtop = \gevcc{\measStatSyst{170.5}{2.4}{1.2}} \cite{d0lj}.

\section{\label{sec:future}Future prospects}
As Run II progresses at the Tevatron, top quark mass measurements are rapidly approaching systematic limits. A new set of analyses are emerging from the Tevatron that make very different assumptions to measure the top quark mass, and as such are sensitive to very different systematic uncertainties. In one such measurement, D\O\  measures the top quark mass via a measurement of the \ttbar pair production cross section. Top quark pairs at the Tevatron are produced nearly at threshold, so the cross section depends strongly on the top quark mass. The analysis depends on theoretical inputs to model this relationship; using \invfb{0.9} of data \cite{d0xsec}, D\O\  measures \mtop = \gevcc{\measStatSystDZero{166.9}{5.9}{5.2}{3.7}{3.8}} using a $\sigma_{\ttbar}$-$\mtop$ curve from Kidonakis and Vogt. Using a curve from Cacciari et al. gives \mtop = \gevcc{\measStatSystDZero{166.1}{6.1}{5.3}{4.9}{6.7}}.

\begin{figure}[htbp]
\vspace{7.0cm}
\includegraphics{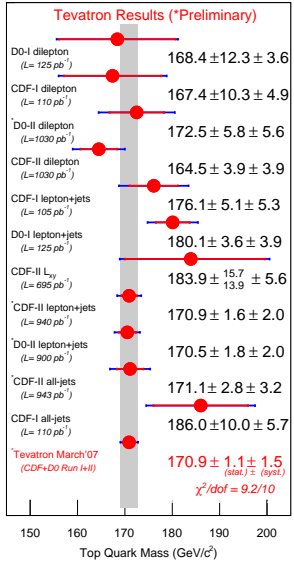}
\caption{\it World-average top quark mass measurement and comparison with individual measurements from the two Tevatron experiments.}
\label{tevplot}
\end{figure}

\begin{figure}[htbp]
\vspace{7.0cm}
\includegraphics{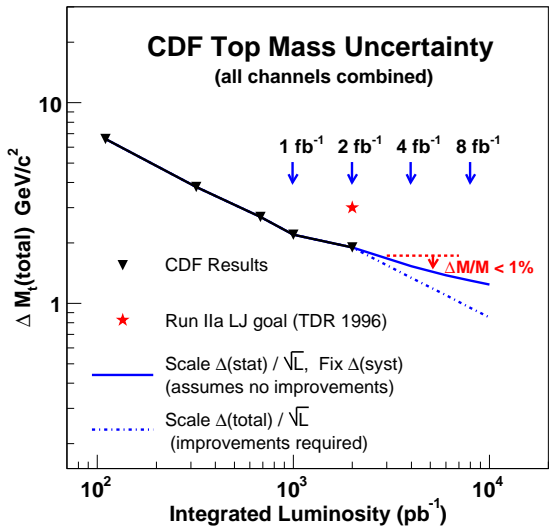}
\caption{\it Projected uncertainty on the top quark mass from CDF-only combinations. The y-axis is the total uncertainty on the top quark mass, combining the statistical and systematic uncertainties in quadrature. The solid blue line is a projection where only statistical uncertainties scale as the inverse square root of the luminosity. The dotted blue line is a projection where both statistical and systematic uncertainties are scaled.}
\label{cdfplot}
\end{figure}

The world average Tevatron top quark mass from the Tevatron as of March 2007, \mtop = \gevcc{\measStatSyst{170.9}{1.1}{1.5}} \cite{cdftev}, already comes close to being a 1\% measurement, and does not include most of the analyses describe in this letter. Figure \ref{tevplot} compares the world average with measurements from both experiments. CDF has a new combination of its own analyses, yielding \mtop = \gevcc{\measStatSyst{172.9}{1.2}{1.5}} \cite{cdfcomb}, and expects by the end of Run II to have a CDF-only combination of top quark mass measurements with a precision better than 1\%, as indicated in Figure \ref{cdfplot}.

\section{Acknowledgements}
I would like to thank all those in the CDF and D\O\ top groups who came before me in Run I and Run II (including all conveners) for setting the bar so high for top quark mass measurements. In addition, many thanks to the organizers of the La Thuile 2008 conference for all their hard work.

\end{document}